\documentclass[aps,pra,twocolumn,
superscriptaddress,showpacs]{revtex4-2}
\usepackage{graphicx}

\begin{document}
\title{ 
Measuring the acceleration of an elevator by using  the apparent weight of an object inside it}
\thanks{This is English translation of a Chinese article published in  {\em Physics Teaching} (Chinese)  {\bf 44} (5), 71-77,80 (2022).} 

\author{Mingyuan Shi}

\affiliation{The Second School Affiliated to Fudan University,  Shanghai 200438, China
} 

\author{Yu Shi}

\affiliation{
Department of Physics, Fudan University, Shanghai 200438, China}


\begin{abstract}
 An accelerating elevator changes the apparent weight of any object inside it from the original weight,  as measured inside  the elevator, because  the acceleration causes an inertial force on it. For any object in a running elevator, the variation of the acceleration of the elevator causes the variation of the apparent weight of the object. We have studied the time dependence of the apparent weight of the object and thus the acceleration of the elevator. For chosen initial and final floors, we measured the apparent weight of an object by using an electronic scale inside the elevator, and shot the readings of the scale and a watch during the movement of the elevator.  Then we analyzed the data collected from the recorded video.   If the initial and final floors are exchanged,  the variations of the weight and acceleration are, respectively, same in magnitudes and opposite in signs.  The experiments indicate that for the elevator to go directly from a floor to another, the process consists of periods with variable acceleration, constant acceleration, uniform motion, variable deceleration, constant deceleration and variable deceleration consecutively.  If there are pauses during the movement,   each pause   causes an additional process  consisting of periods  with deceleration, stop and acceleration, replacing the original period of constant motion. Depending on the distance to the destination, the elevator reduces or   diminishes the periods of constant acceleration and uniform motion.    
\end{abstract}

\maketitle 
\section{Introduction}

The reference frame of a freely falling object is a non-inertial frame, the acceleration of which is equal to the gravitational acceleration, which is defined as the weight $\mathbf{W}$ of any object on the surface of the earth divided by its mass $m$.  In this frame, the freely falling object itself is static, and is also subject to an inertial force, which is a fictitious force  with its magnitude same as  and its direction  opposite  to  those of weight and thus cancels the weight.  If, instead of freely falling, a person accelerates with a different acceleration, she or he also feels a change of weight, either a loss  or a gain of the apparent weight, which is equal to the original weight adds the inertial force, which is equal to minus the acceleration multiplied by the mass.  Another example of the inertial force is the centrifugal force felt by an object in uniform circular motion. If the  rotation of the earth is taken into  account, the earth itself becomes a non-inertial frame, and the  weight is the gravitational force  deducted by the centrifugal force  and  Coriolis force.   In 1907, having realized that a freely falling person does not feel her or his own gravity, Einstein proposed that the gravity is equivalent to the acceleration, and is itself an inertial force, and finally founded the general theory of relativity in 1915. 

In this paper, we measure the acceleration of a running elevator by using the apparent weight of an object inside the elevator. This not only demonstrates the nature of the   apparent weight and the inertial force, but also displays the  consecutive periods of the motion of the elevator, including accelerated motion, uniform motion, and decelerated motion. The accelerated motion consists of periods  with the increasing acceleration, with   constant acceleration,  and with decreasing acceleration. The  decelerated motion consists of   periods  with increasing magnitude of acceleration, with constant acceleration, and with decreasing magnitude of acceleration.  During the accelerated motion, the acceleration  is in the same direction as the velocity,  it  first increases, then remains  maximal as the constant acceleration,  then decreases. Subsequently, the elevator undergoes uniform motion. During the decelerated motion, the acceleration is in the opposite direction to the velocity,  its magnitude first increases, then remains maxium, and then decreases.  Subsequently, the elevator stops. If the initial and final floors exchange, the variation of the acceleration almost remains the same in magnitude, but opposite in direction. 

If there are pauses in the process, each pause  causes an additional process  consisting of periods of decelerated motion,  pause and accelerated motion, replacing the original process of  uniform  motion. Depending on the distance to the final stop, the elevator reduces or even diminishes the periods of  constant acceleration and  of uniform motion.    

\section{Principle}
 
 A non-inertial reference frame, also called accelerated reference frame,  is one with acceleration $\mathbf{a}$  itself, with respect to an inertial frame.  Consider an object with mass $m$. Suppose its acceleration is $\mathbf{a}_1$  with respect to this non-inertial frame. Then its acceleration with respect to the inertial frame is  $\mathbf{a}+\mathbf{a}_1$ . 
 
 In the inertial frame, Newton's Second Law gives rise to 
 \begin{equation}
 \mathbf{F} = m(\mathbf{a}+\mathbf{a}_1), \label{f}
 \end{equation}
 where  $\mathbf{F}$ is the force acting on it.  (\ref{f}) can be rewritten as 
  \begin{equation}
 \mathbf{F}+\mathbf{F}_1 = m\mathbf{a}_1, \label{f1} 
 \end{equation}
 where
  \begin{equation}
  \mathbf{F}_1 = -m\mathbf{a}
  \end{equation}
  is the inertial force~\cite{kittle}.   Therefore in the accelerated frame, for  the acceleration $\mathbf{a}_1$ to satisfy the Newton's second law in the appearance, there exists a fictitious force  $\mathbf{F}_1=-m\mathbf{a}$,  in addition to the real force  $\mathbf{F}$,  e.g. the weight. In considering the rotation of the earth, the weight  itself already includes the inertial force due to the  rotation of the earth in addition to the gravitational force. 
  
  Consider the weight of an object   
   \begin{equation}
  \mathbf{W} = m \mathbf{G},
  \end{equation}
 where $\mathbf{G}$ is the gravitational acceleration, with the magnitude $G \approx 9.8 m/s^2$.  
 
 When the object rests on the earth or  in a static elevator, its apparent weight $\mathbf{W}'$ reduces  to the weight   $\mathbf{W}$. This is because the pressing force on the scale is equal in magnitude and opposite in direction   to the normal (supporting) force  $\mathbf{N}$, which  is equal in magnitude and opposite in direction to the weight  $\mathbf{W}$, hence  the pressing force on the scale is equal to   $\mathbf{W}$.  That is,  now $\mathbf{W}'=-\mathbf{N}=\mathbf{W}$.   The unit of the reading of the scale is  re-scaled to that of the mass, that is, the measured quantity is actually  $\mathbf{W}'$  while the reading is   $W'/G$.  When the elevator is static, $W'/G$ is the mass $m$. 
 
 When the elevator accelerates, in the reference frame of the elevator,  the object inside it is subject to the inertial force  $-m\mathbf{a}$, in addition to the weight $\mathbf{W}$. In the reference frame of the earth, it is because the pressing force is not equal to the weight that  the object accelerates together with the elevator. Therefore, the apparent weight of $W'$ is 
 \begin{equation}
 \mathbf{W}'=\mathbf{W}-m\mathbf{a}. 
 \end{equation}
   Correspondingly,   the reading of the scale gives  $m' \equiv W'/G=m(1-a/G)$. 
 
 Note that $m'$ is not a real mass, and is  $W'/G$, i.e. the apparent weight re-scaled in terms of the unit of mass. For convenience, sometimes we simply refer to $m'$ as the apparent weight. 
 
 Now in the accelerated frame,  Eq.~(\ref{f1}) reads 
 \begin{equation}
 \mathbf{W}+\mathbf{N}-m\mathbf{a}=0,
 \end{equation}
  while 
 \begin{equation}
  \mathbf{N}=- \mathbf{W}', 
 \end{equation}
  hence  one again obtains  $\mathbf{W}'=\mathbf{W}-m\mathbf{a}$. Hence 
 \begin{equation}
 \mathbf{a} =(1-\frac{m'}{m})\mathbf{G}.
 \end{equation}
 
 When $m'<m$,  the acceleration of the elevator is along  the same direction as the gravitational acceleration, this is the case of weight loss; when  $m'>m$,  the acceleration of the elevator is opposite to the gravitational acceleration, this is the case of overweight. 
 
In this paper, we use positive number for the acceleration  $\mathbf{a}$  if it is along the   direction  of the gravitational acceleration, while use negative number if it is opposite to the direction of  the gravitational acceleration. 
 
 We first measure $m'$ during the movement of the elevator, then  obtain $\mathbf{a}$.

\section{Experiment}

 We first found a suitable elevator. 
 The experimental process is as follows. 
 \begin{enumerate}
 \item Prepare the scale and the object. We use a  heavy book, together with a  mobile phone to be used as the watch,  as the object, and measure its weight. When the elevator is static, the reading for the  weight of the book was  found to be $m=4.10kg$. 
 \item Put the scale inside the elevator. The object, i.e. the book  together with a watch, is put on the scale.  
\item  Record a video about the readings of scale and the watch during the movement of the elevator. 
\item  For  each  movement, from an initial level to a final level,  make  the video recording, and repeat for three to four times. 
\end{enumerate}

 For each process from an initial level to a final level,  the experimental data are collected from the video record in the following way.  
 \begin{enumerate}
 \item   Note down the starting time.  
 \item   For arbitrarily chosen times with intervals about $0.5s$ to $1s$,  pause  the video, note down the times and readings of  $m'$. The time differences with the starting time is defined as $t$. 
 \item   Plot $m'$ and   the acceleration $a$ as  functions of $t$.   The unit of   $m'$  is $kg$, and the unit of $a$ is $G$. The unit of time is second ($s$). 
\end{enumerate}

Using the same method, we study the processes  that the elevator runs from Level B1 to Level 4, that  from Level 4 to Level B1,  that from Level B1 to Level 3, and that from Level 3 to Level B1. We also study the situation that the elevator pauses and restarts during the processes  that  from Level  B1 to Level 2 and that from Level B1 to Level 4. Each process is  repeated for 3 or 4 times. 

\section{Data Analysis} 

\begin{figure}[b]
\centerline{\includegraphics[width=10cm]{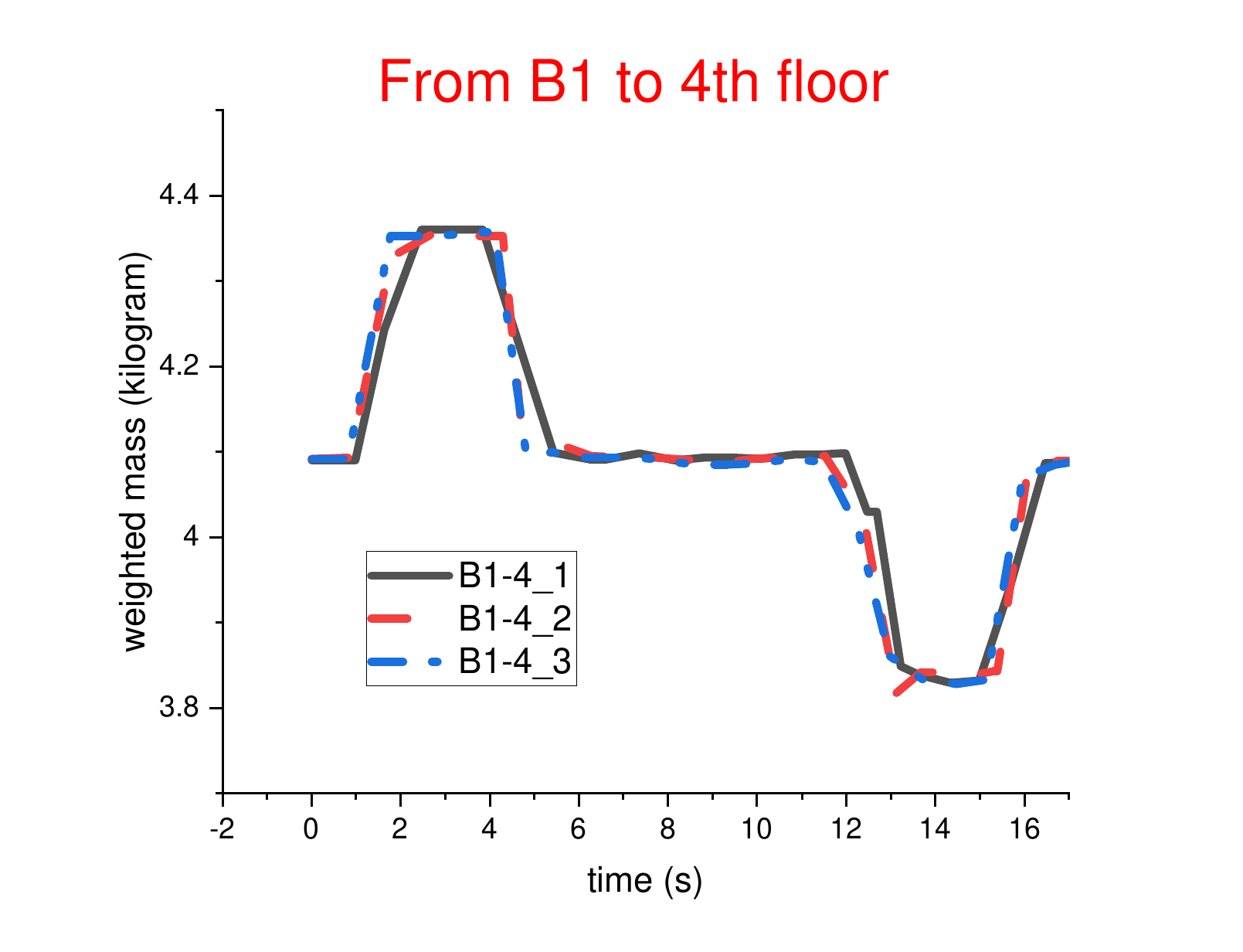}}
\centerline{\includegraphics[width=10cm]{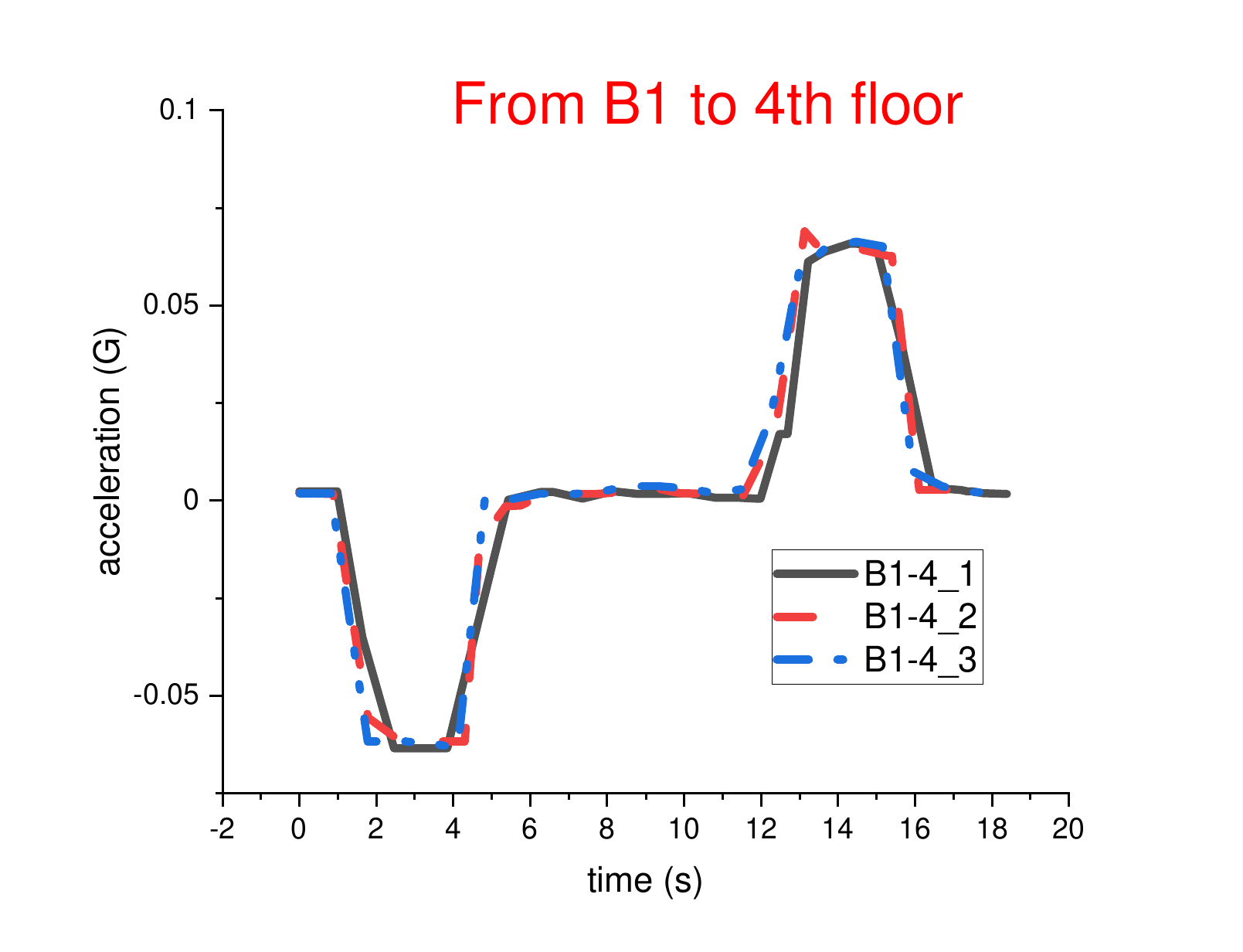}}
\caption{(a) The dependence of the apparent weight on time,   
(b) The dependence of the acceleration of the elevator on time, when the elevator runs from Level B1 to Level 4.    \label{fig1}}
\end{figure}

\begin{figure}[b]
\centerline{\includegraphics[width=10cm]{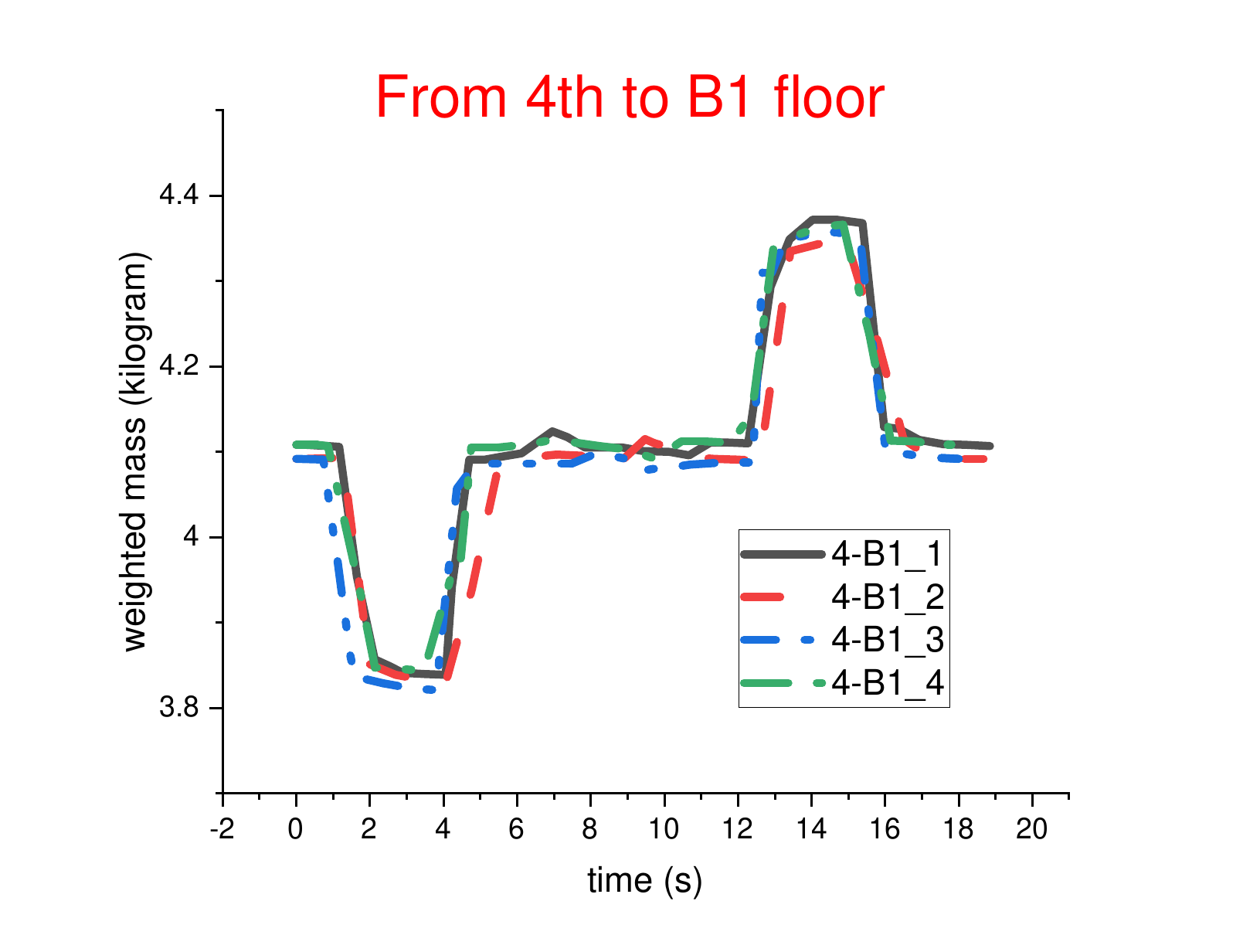}}
\centerline{\includegraphics[width=10cm]{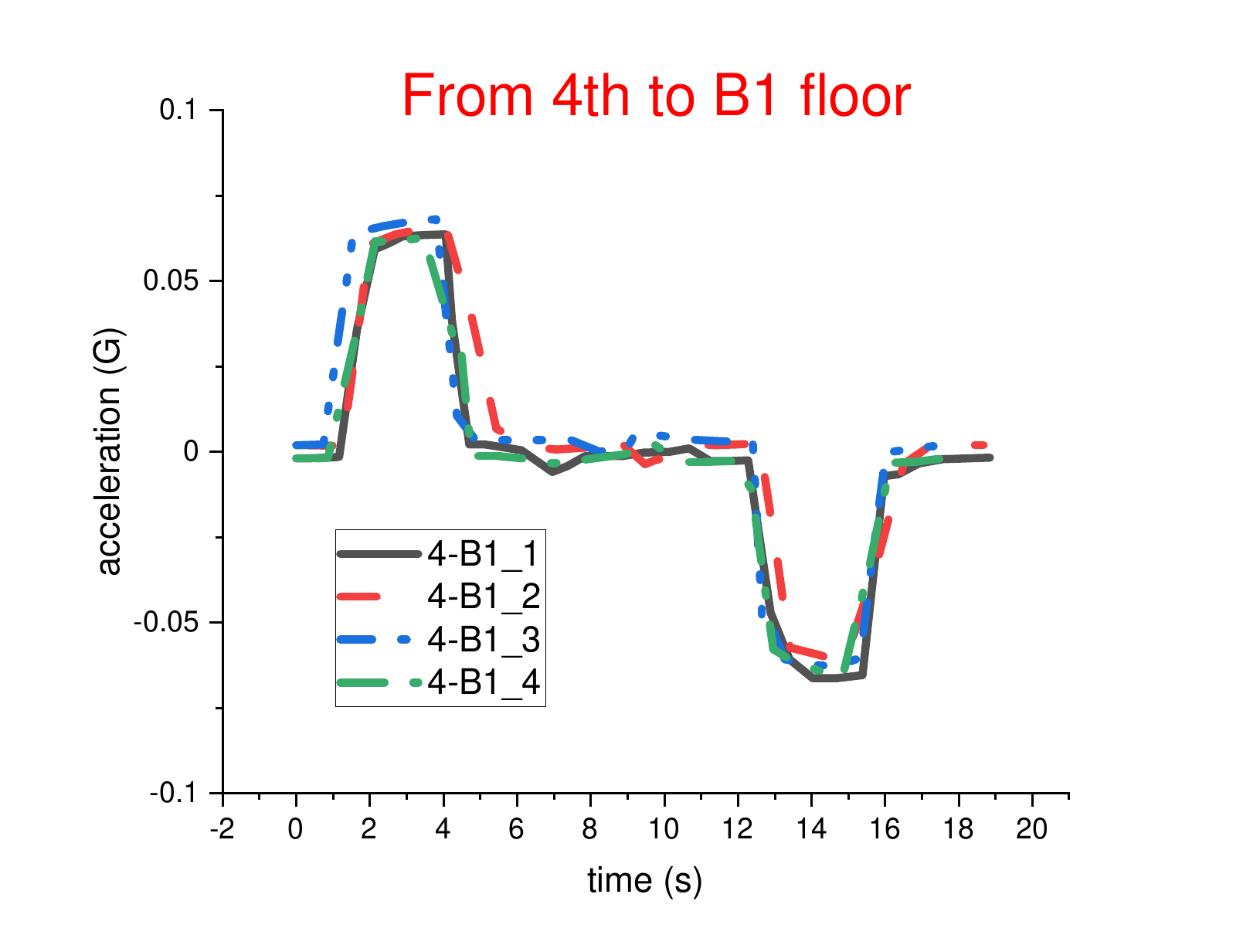}}
\caption{(a) The dependence of the apparent weight on time,  
(b) The dependence of the acceleration of the elevator on time, when the elevator runs from Level 4  to Level B1.    \label{fig2}}
\end{figure}

\begin{figure}[b]
\centerline{\includegraphics[width=10cm]{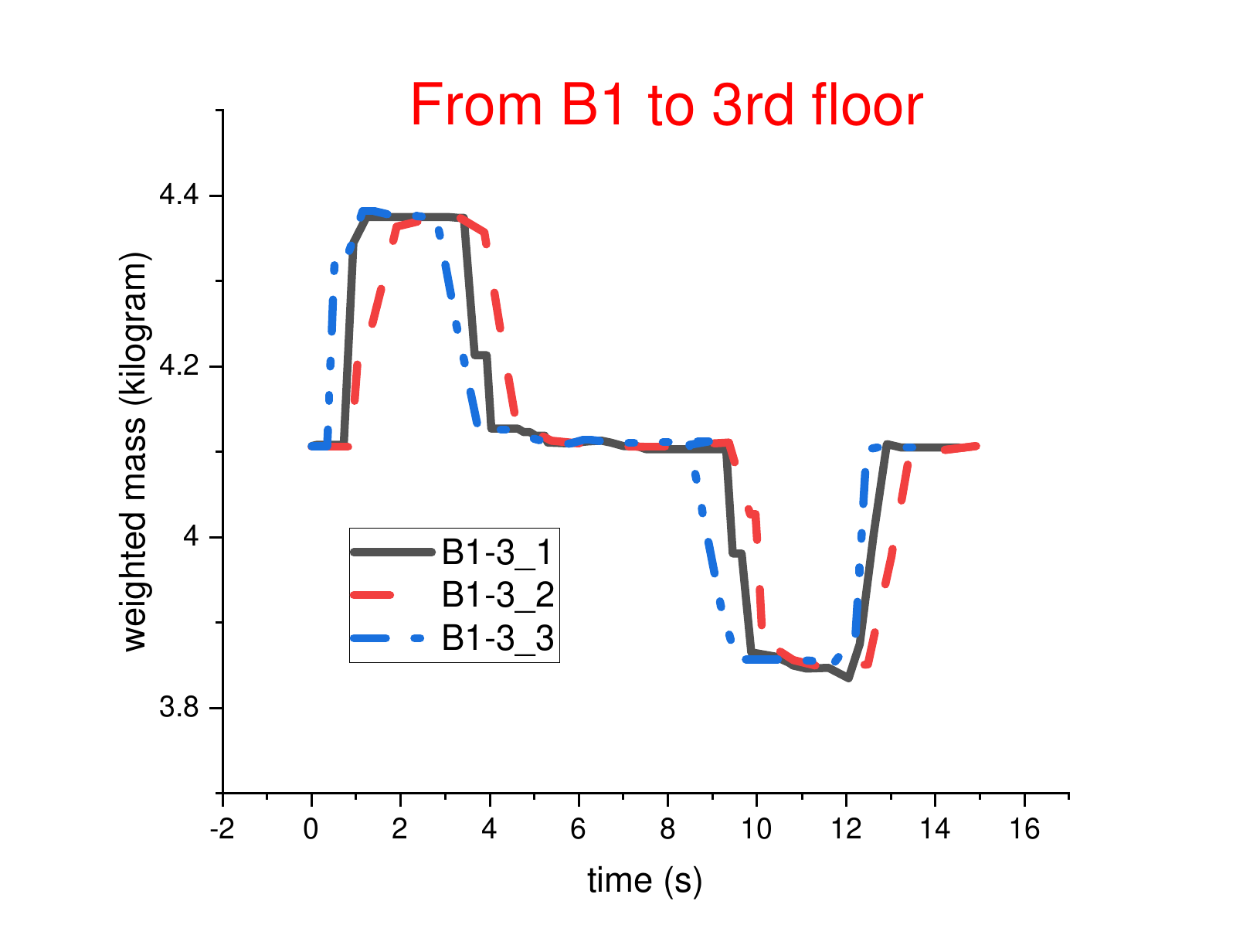}}
\centerline{\includegraphics[width=10cm]{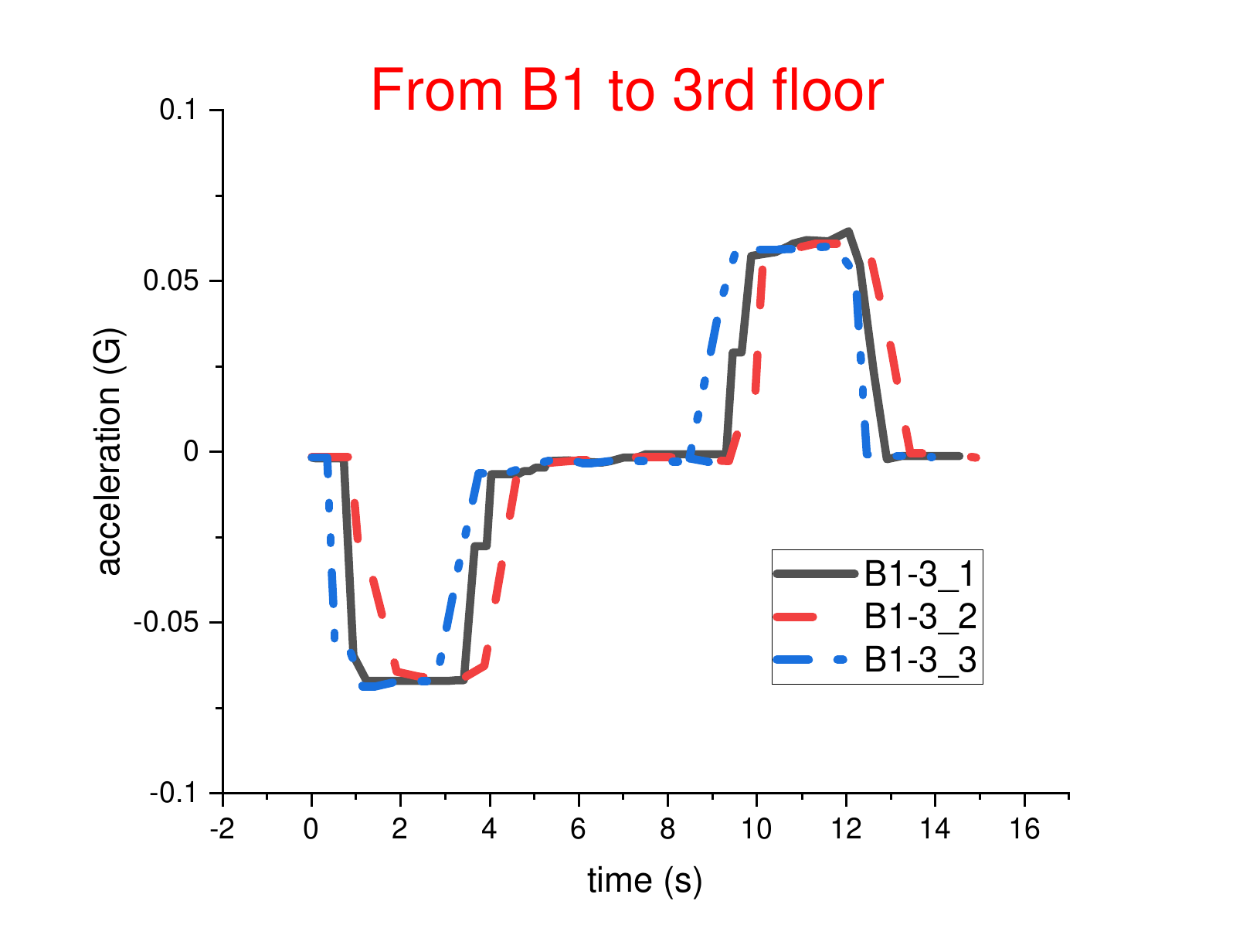}}
\caption{(a) The dependence of the apparent weight on time,  
(b) The dependence of the acceleration of the elevator on time, when the elevator runs from Level B1 to Level 3.    \label{fig3}}
\end{figure}

\begin{figure}[b]
\centerline{\includegraphics[width=10cm]{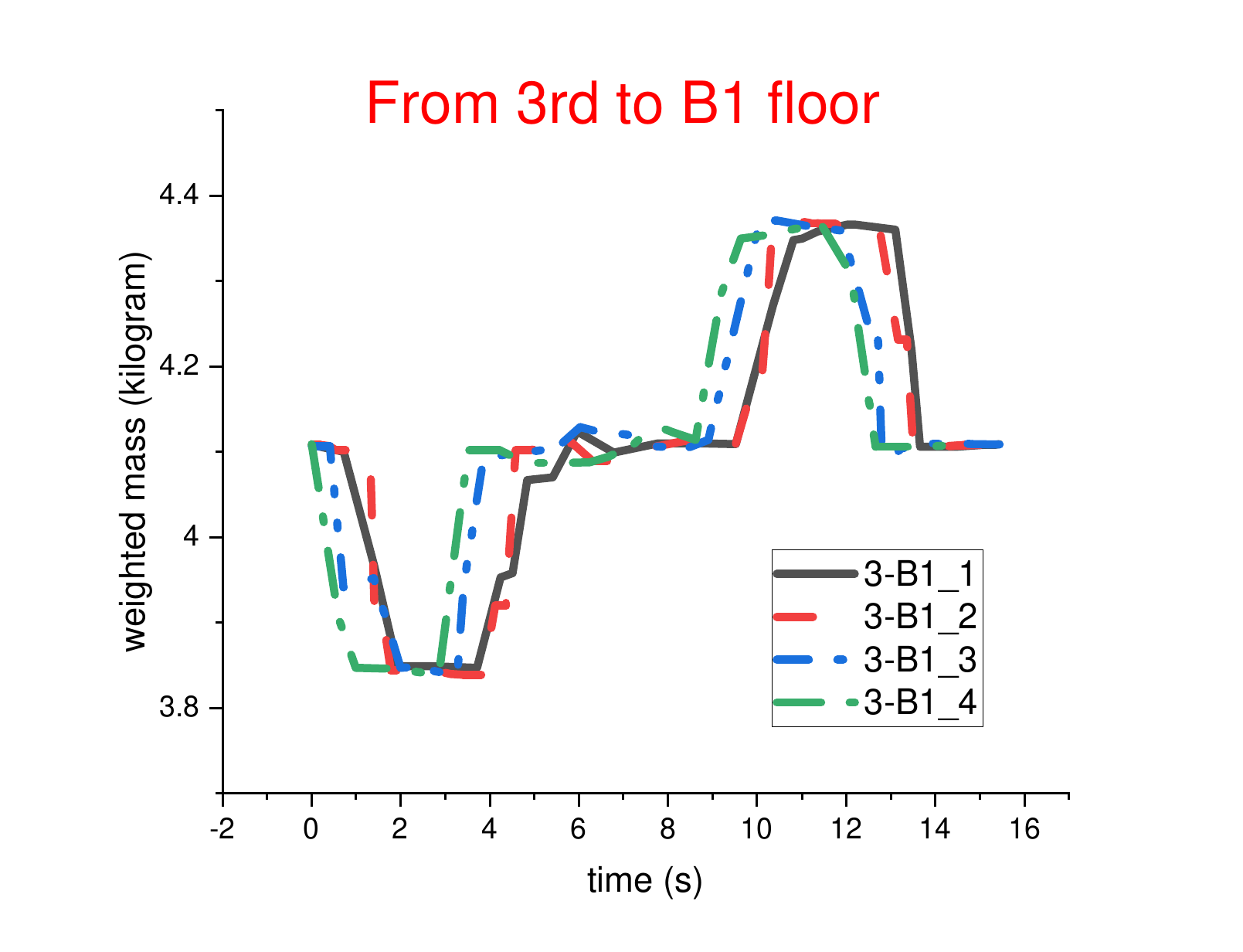}}
\centerline{\includegraphics[width=10cm]{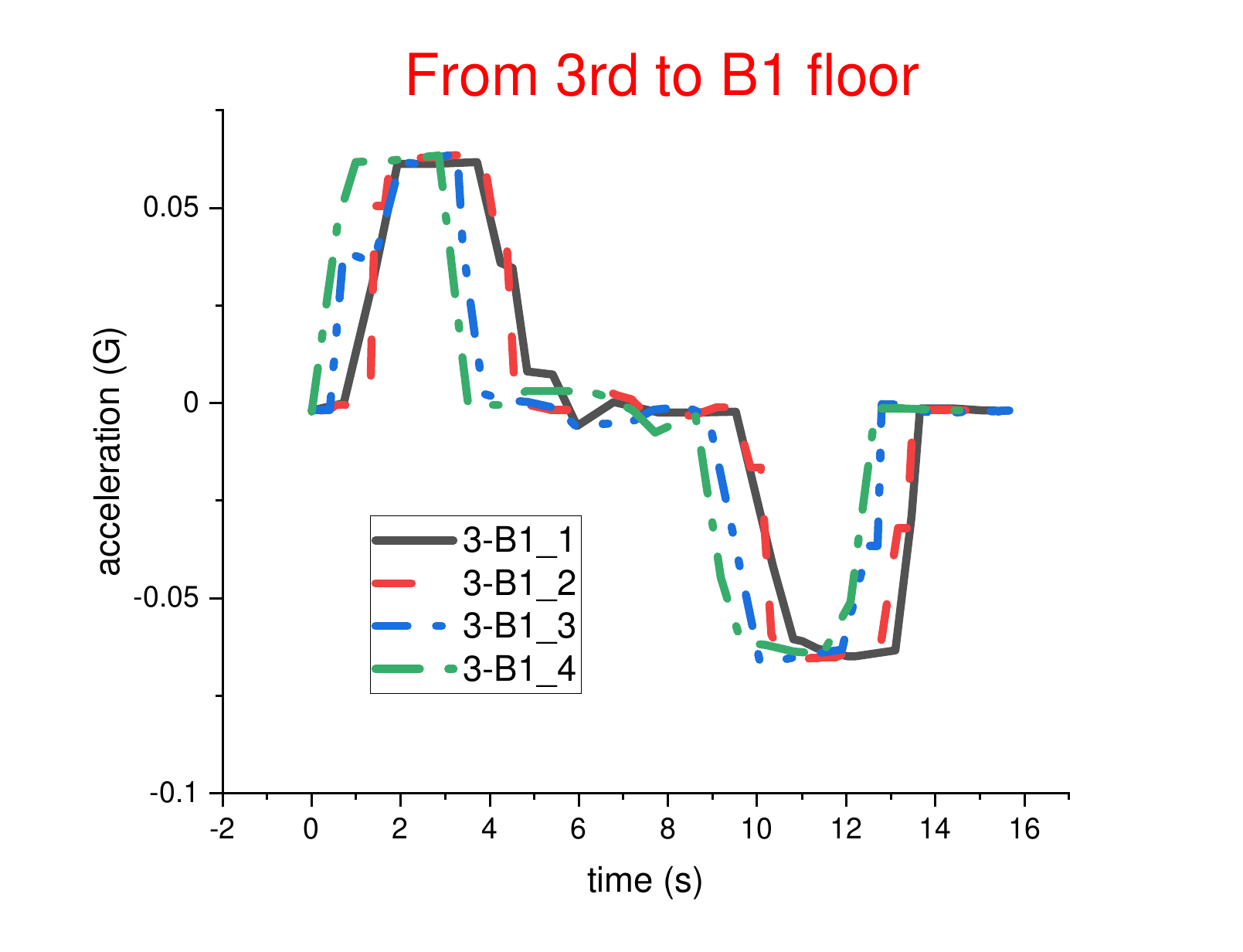}}
\caption{(a) The dependence of the apparent weight on time,  
(b) The dependence of the acceleration of the elevator on time, when the elevator runs from Level 3 to Level B1.    \label{fig4}}
\end{figure}

\begin{figure}[b]
\centerline{\includegraphics[width=10cm]{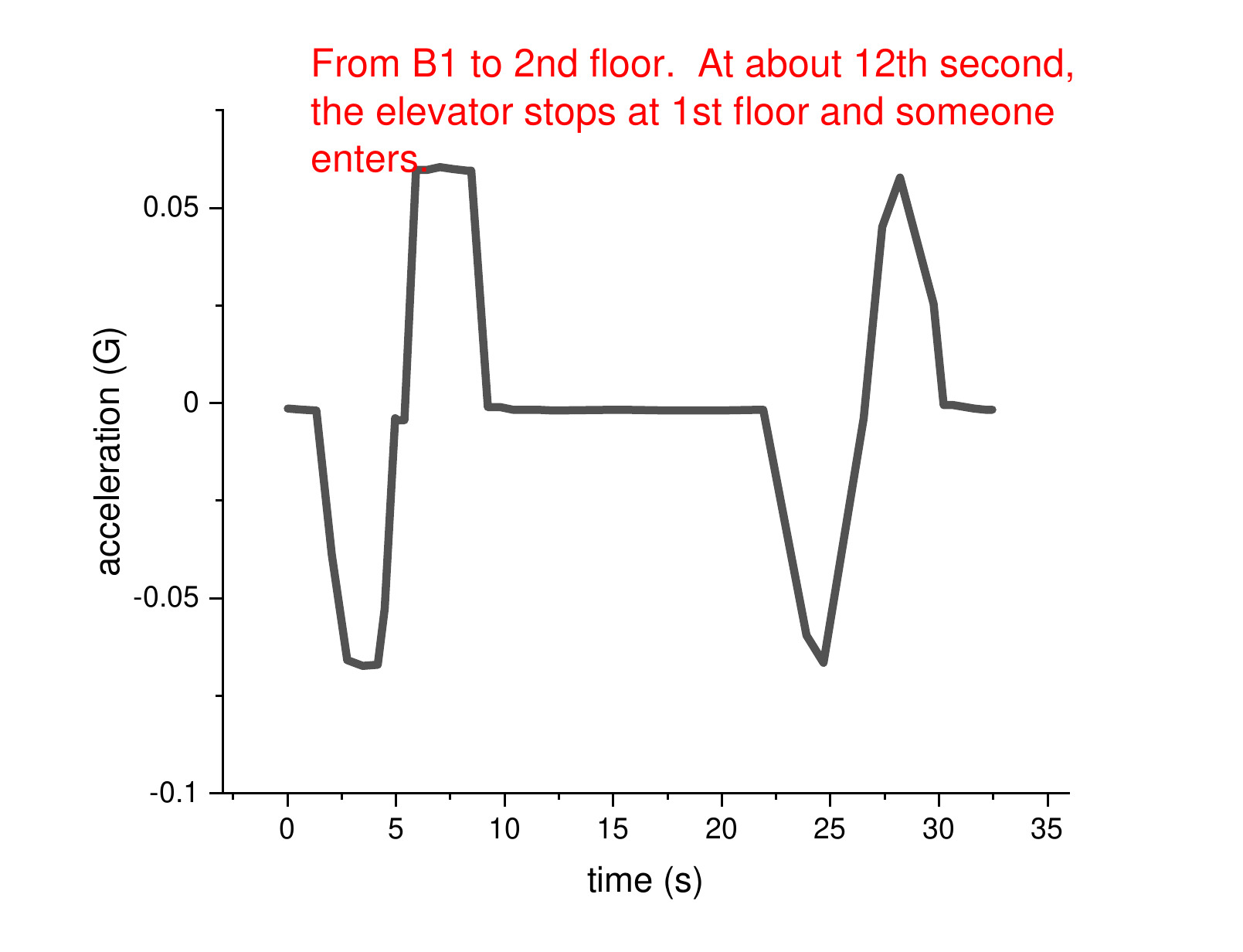}}
\centerline{\includegraphics[width=10cm]{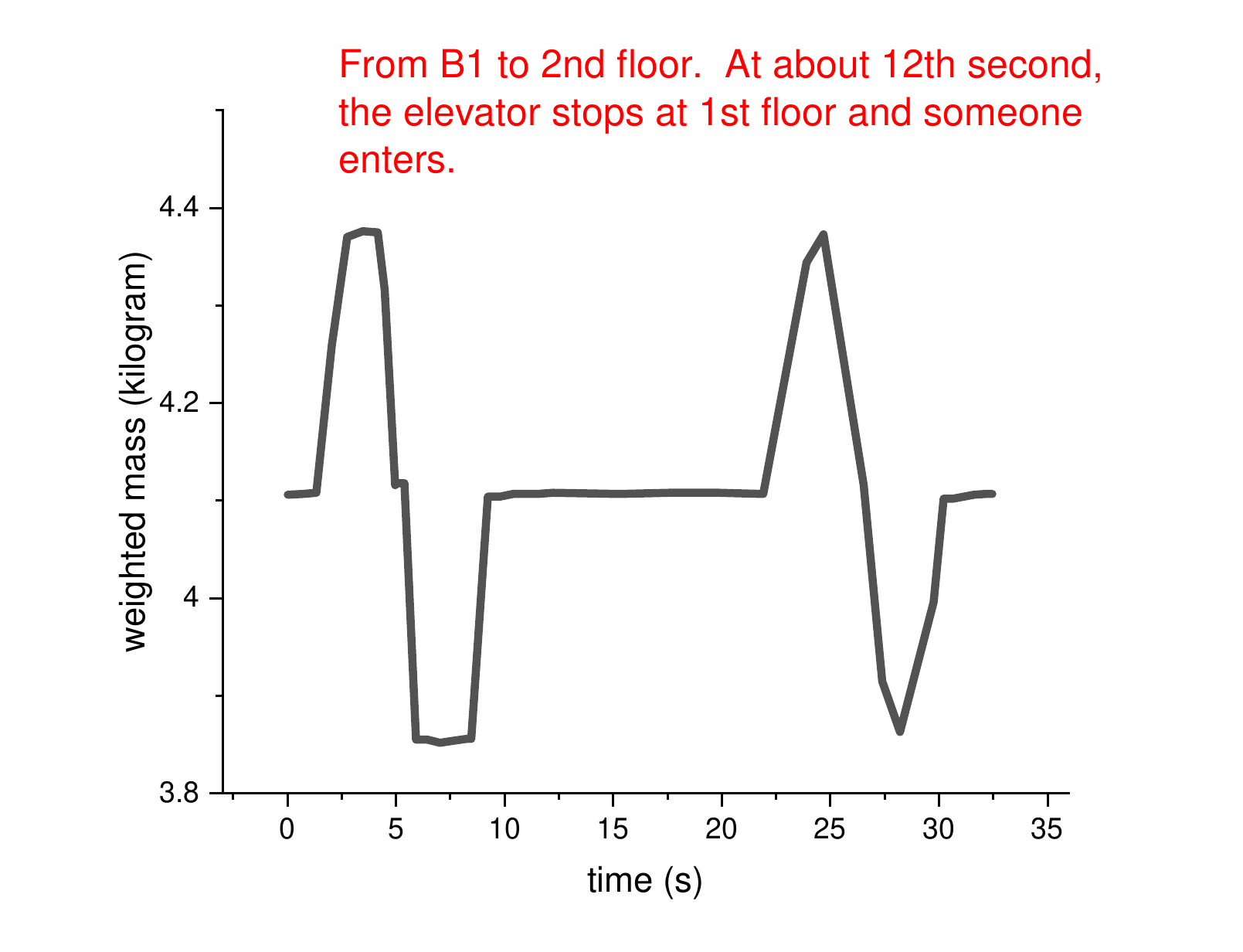}}
\caption{(a) The dependence of the apparent weight on time,  
(b) The dependence of the acceleration of the elevator on time, when the elevator runs from Level B1 to Level 2, with a pause on Level 1.    \label{fig5}}
\end{figure}

\begin{figure}[b]
\centerline{\includegraphics[width=10cm]{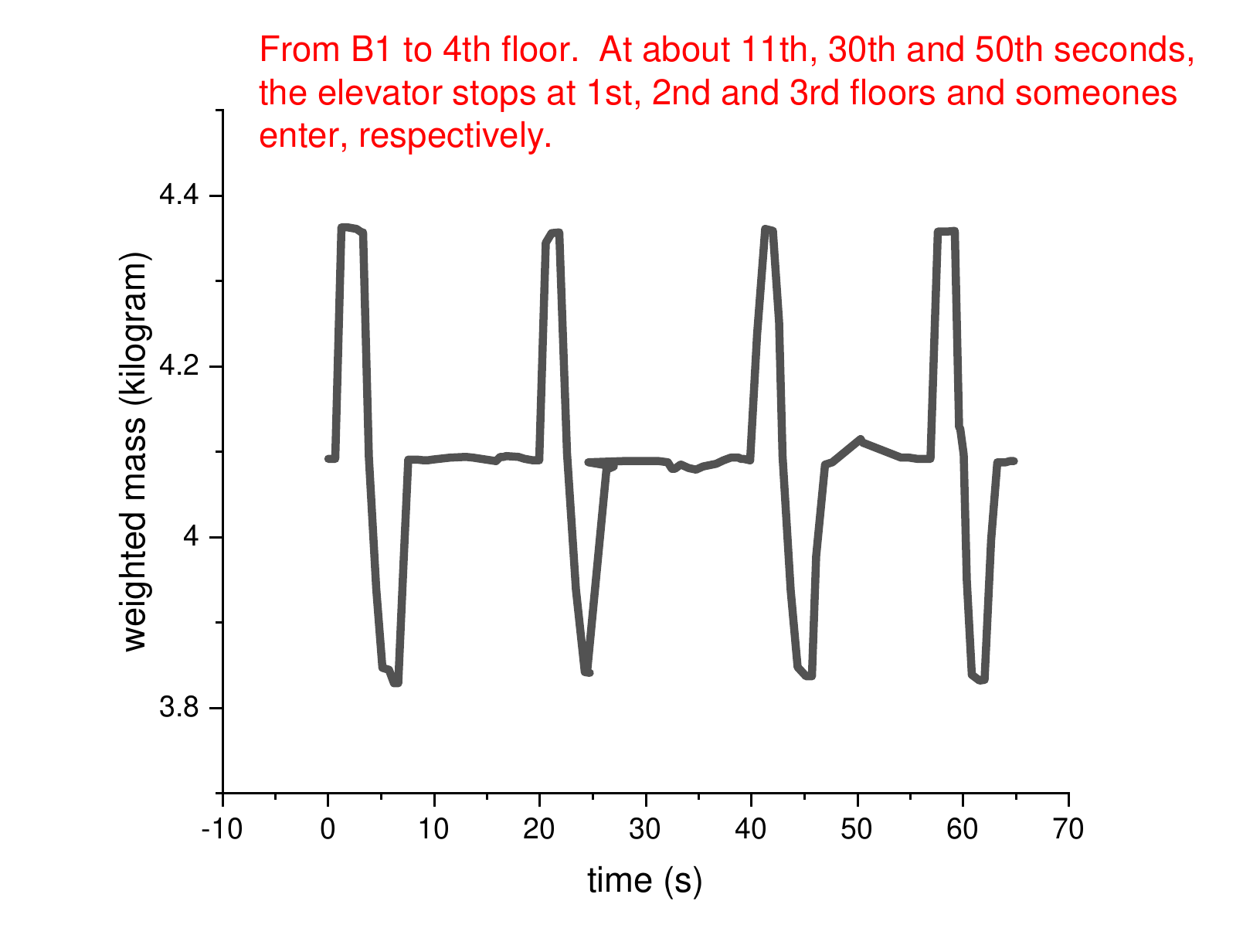}}
\centerline{\includegraphics[width=10cm]{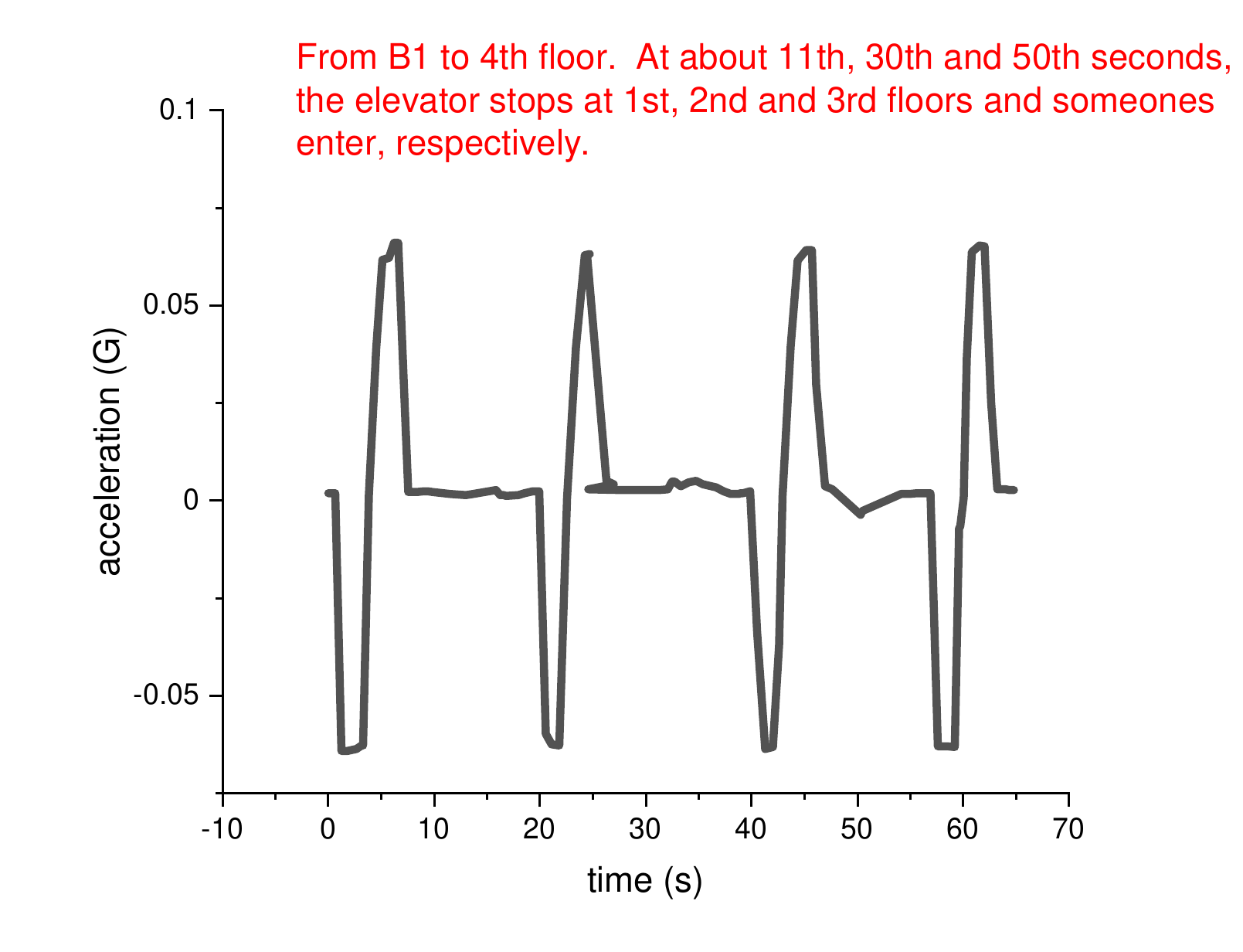}}
\caption{(a) The dependence of the apparent weight on time,  
(b) The dependence of the acceleration of the elevator on time, when the elevator runs from Level B1 to Level 4, with pauses on Levels 1, 2 and 3.    \label{fig6}}
\end{figure}

For each process,  we obtain the dependence of $m'$ on $t$   and that of $a$  on $t$, as shown in Fig. (\ref{fig1}) to Fig. (\ref{fig6}) . For  the processes investigated in  Fig. (\ref{fig1}) to (\ref{fig5}), about $40$ data points are collected for each. For the processes investigated in  Fig. (\ref{fig6}) , about $100$ data points are collected for each. 
  
\subsection{Non-stop movement  from the initial to the final floors  }

Fig. (\ref{fig1}) indicates that during the process from Level B1 directly  to Level 4, how $m'$ and $a$ depend on time.  Three sets of data, from the three times of  repeating this process, are obtained, which are largely similar, indicating that these processes are mainly determined by the intrinsic properties of the elevator. The minor differences are mainly due to reason that  the initial time is not exactly certain, as the apparent weight of the object inside the elevator is almost unchanged during the short period after the door is closed,   after which the elevator starts to  move.  

In this process, the elevator moves upwards, thus the velocity is always upwards. When the acceleration is also upwards, the speed increases. When the acceleration  is downwards, the speed decreases. In our convention, the acceleration is negative when it is  upwards,  while it is positive when it is downwards.  

From Fig. ~\ref{fig1}, it can be seen that the whole process consists of the following $7$ periods. 
\begin{enumerate}  
\item An accelerating period with an increasing magnitude of acceleration. As shown in Fig. 1a, at about $1s$, the elevator starts to move, and   $m'$ starts to increase from $m$, and  as shown in Fig.~\ref{fig1}b,  the magnitude of the acceleration starts to increase, and thus  the speed increases.  This is the case of  overweight, and the acceleration  is negative. 
\item An accelerating period with a constant acceleration.  At about $2s$, the apparent weight reaches the maximum, correspondingly, the magnitude of the acceleration of the elevator  reaches the maximum. As seen in Fig.~1, the apparent weight and the magnitude of  the acceleration keep the maxima respectively. The speed keeps increasing. 
\item An accelerating period with a decreasing magnitude of acceleration. As shown in Fig.~1,  at about $4s$, the apparent weight $m'$ starts to decrease,  but remains larger than $m$, so the  overweight still goes on. The magnitude of the acceleration, which is negative, starts to decrease, but its direction is still the same as that of the velocity. hence the speed remains increasing. So the elevator keeps going upwards. 
\item A period of uniform motion. At about $5s$, $m'$ returns to $m$, and the acceleration goes back to $0$. Before this, the direction of the acceleration remains the same as the direction of the velocity, so the speed remains increasing. So in this period of uniform motion, the velocity is nonzero, and upwards. 
\item A decelerating period with increasing magnitude of  acceleration.  At about $12s$, the apparent weight $m'$ decreases from $m$. so this is weight loss. The acceleration is now positive, i.e. its direction is downwards and the magnitude keeps increasing. As the velocity is upwards, opposite to that of the acceleration, the speed decreases. The magnitude of the acceleration remains increasing, while the speed  decreases. 
\item   A decelerating period with constant acceleration. At about $13s$, the apparent weight reaches its  minimum, correspondingly the magnitude of the  acceleration  reaches its maximum. As seen from Fig.~\ref{fig1}, for a while, the apparent weight remains the minimum, and the  magnitude of the acceleration remains the maximum. The velocity is upwards, opposite to the acceleration, so the speed keeps decreasing. 
\item  A decelerating period with decreasing magnitude of acceleration. At about $15s$, the apparent weight $m'$ starts to increase, but is still less than $m$, so there is still weight loss. The acceleration, which is positive, starts to decrease in magnitude, but the direction remains opposite to the velocity, which is upwards. Hence the speed keeps decreasing. At about $16s$, $m'=m$, the acceleration becomes $0$, from which one cannot determine the velocity. However, the decelerating process from $12s$  to $16s$ is symmetric with the accelerating process from $1s$  to $5s$, indicating that the deceleration cancels the acceleration. In fact, now the elevator arrives at the destination, and the velocity must be $0$. 
      \end{enumerate}
      
In the above detailed analysis, the first three periods  consist of an accelerating process, with the acceleration in the same direction of the velocity, and the elevator leaving the initial place; the fourth period is a process of uniform motion, with the acceleration being 0, and the duration  dependent on  the distance between the initial floor and the destination, the longer the distance, the longer this period;  the last three periods consist of a decelerating process, with the acceleration opposite to the velocity, and the elevator approaching the destination. 
      
For the reversal process, from Level 4 to Level B1,  how the apparent weight and the acceleration depend on time is shown in Fig.~\ref{fig2}. In this process, the elevator moves downwards, so the velocity is always downwards. When the acceleration is also downwards, the speed keeps increasing; when the acceleration is upwards, the speed keeps decreasing.  The whole process   also consists of $7$  periods. 
   
\begin{enumerate}  
\item An accelerating period with an increasing acceleration. As shown in Fig.~\ref{fig2}a, at about $1s$, the elevator starts to move, and   $m'$ starts to decrease from $m$, and  as shown in Fig.~\ref{fig2}b,  the  acceleration starts to increase, and thus  the speed increases.  This is the case of  weight loss, and the acceleration  is positive. 
\item An accelerating period with a constant acceleration.  At about $2s$, the apparent weight reaches the minimum, correspondingly, the acceleration of the elevator  reaches its maximum. As seen in Fig.~\ref{fig2},  the apparent weight keeps its minimum and the acceleration keeps  its  maximum, respectively. The speed keeps increasing. 
\item An accelerating period with a decreasing acceleration. As shown in Fig.~2,  at about $4s$, the apparent weight $m'$ starts to increase,  but remains less  than $m$, so the  loss of weight still goes on. The  acceleration starts to decrease, but its direction is still the same as that of the velocity. hence the speed remains increasing, and  keeps downwards. 
\item A period of uniform motion. At about 5th second, $m'$ returns to $m$, and the acceleration returns to $0$. Before this, the direction of the acceleration remains the same as the direction of the velocity, so the speed remains increasing. So in this period of uniform motion, the velocity is nonzero and downwards. 
\item A decelerating period with increasing magnitude of  acceleration.  At about $12s$, the apparent weight $m'$ increases from $m$. so this is overweight. Correspondingly, the acceleration becomes negative, i.e. its direction is upwards and keeps increasing in magnitude. The velocity is downwards, opposite to that of the acceleration, so the speed decreases. The magnitude of the acceleration remains increasing, while the speed  decreases. 
\item   A decelerating period with constant acceleration. At about $13s$, the apparent weight reaches its  maximum, correspondingly the acceleration of the elevator reaches its maximum. As seen from Fig.~\ref{fig2}, for a while, the apparent weight remains the maximum, and the magnitude of the acceleration remains the maximum. The velocity is downwards, opposite to the acceleration, so the speed keeps decreasing. 
    \item  A decelerating period with decreasing magnitude of acceleration. At about $15s$, the apparent weight $m'$ starts to decrease, but is still larger than $m$, so there is still overweight. The acceleration, which is negative, starts to decrease in magnitude, but the direction remains opposite to the velocity,  hence the speed keeps decreasing. At about $16s$, $m'=m$, the acceleration becomes $0$, from which one cannot determine the velocity. However, the decelerating process from $12s$ to $16s$ is symmetric with the accelerating process from $1s$ to $5s$, indicating that the deceleration cancels the acceleration. In fact, now the elevator arrives at the destination, and the velocity must be $0$. 
      \end{enumerate}
   
Similar to the process from Level B1 to Level 4,  in the process from Level 4 to Level B1,   the first three periods comprise  an accelerating process, with the acceleration in the same direction as the velocity, and the elevator leaving the initial place; the fourth period is a process of uniform motion, with the acceleration being $0$, and the duration  determined by the distance between the initial place and the destination; and the last three periods comprise  a decelerating process, with the acceleration opposite to the velocity, and the elevator approaching the destination. 
   
The two reversal processes, that  from Level B1 to Level 4, as shown in Fig.~\ref{fig1},  and that from Level 4 to Level B1, as shown in Fig.~\ref{fig2}, are symmetric to each other. It can be seen that for each reversal pair of the corresponding periods, the velocity directions and are opposite, and the acceleration directions are also opposite. 
   
We also study  the process from Level   B1 to Level 3. The dependence of the weight and acceleration on time is shown  in Fig.~\ref{fig3}, similar to the process from Level B1 to Level 4.  The reversal process, from Level 3 to Level B1, is also studied. The dependence of the weight and acceleration on time is shown  in Fig.~\ref{fig4}, similar to the process from Level 4 to Level B1.  It can be observed that by switching the initial and final Levels, the velocity and acceleration  both reverse their directions at each instant, and the case of overweight becomes the case of weight loss, and vice versa.  The movement   from one level to another is consisted of  three parts, including acceleration, uniform motion and deceleration. In more details, it  includes seven periods, including the accelerating one with increasing acceleration,       the  accelerating  one with constant acceleration, the  accelerating one with decreasing acceleration, the one of uniform motion, the  decelerating one with increasing magnitude of the deceleration , the   decelerating one constant deceleration and the    decelerating  one with decreasing magnitude of deceleration. 

\subsection{The case with pauses and restarts during the movement  } 
        
We have also studied the case in which the elevator pauses and restarts during its movement. We obtained the times of the pauses and restarts from the data. Shown in Fig.~\ref{fig5} is the running from Level B1 to Level 2, and pauses and restarts on Level 1. After the elevator starts,  the process between  $1s$ and  $5s$ is an accelerating process, including the increase of the magnitude of the acceleration between $1s$  and $2.74s$, a small variation between $2.74s$ and $4.16s$, decrease of the magnitude of the acceleration between $1.16s$  and $5s$.  The acceleration is negative by convention, with the direction upwards, which is the same as the velocity.  Afterwards, there is a period of uniform motion between $5s$ and $5.4s$. This is because the elevator is to pause on Level 1, so the period of uniform motion has to be shortened, and an additional period of deceleration is inserted between $5.4s$ and $5.91s$, in which the acceleration is positive, with the direction  downwards, which is opposite to the direction of the velocity, including the periods of increase in the magnitude of the acceleration between $5.4s$ and  $5.91s$, a very short period  between $5.91s$ and $7.01s$ in which the there is  small variation is  acceleration , and the period of decrease in the magnitude of the acceleration between $7.01s$ and $9.23s$. 

At $9.23s$, the magnitude of the acceleration decreases to 0, from which it is uncertain whether the velocity is $0$. However, according to the symmetry between the deceleration between $5.4s$ and $9.23s$  and  the acceleration between $1s$ and $5s$, their effects cancel, and it may be supposed that the velocity comes back to $0$. Indeed,  now the elevator pauses on Level 1, with the velocity indeed being $0$, and there is someone getting on the elevator. 

At $21.91s$, the elevator restarts, and the acceleration is upwards, negative by convention, while the magnitude keeps increasing. At $23.9s$, the rate of the change of the acceleration decreases in magnitude, and at $24.68s$, the magnitude of the acceleration reaches the maximum.  Now the elevator is close to the destination, the the magnitude of acceleration  ends the period of the  maximum, and tbegins to decrease.  

At  $26.53s$, $m'$ decreases to $m$. and the magnitude of the acceleration decreases to $0$, but there is no uniform motion, as it is very near to the destination, instead, the decelerating period starts. At $27.39s$, the magnitude of the acceleration reaches the maximum, but does not keep it, rather, it  immediately begins to decrease and continue the deceleration till the acceleration becomes 0 at  about $30s$. According to the antisymmetry between the deceleration between $21.93s$  and $26.53s$  and the acceleration between  $26.53s$  and $30s$, their effects cancel. So it can be deduced that the velocity returns to $0$. In fact, now the elevator reaches Level 2, and stops. 

According to the distance away from the destination, the elevator shortens or omits the periods of constant acceleration and of uniform motion. Besides, the elevator inserts the periods of deceleration between $5.4s$ and $9.23s$, the pause  between $9.4s$  and $21.91s$, and the acceleration between $21.91s$  and $26.53s$. These periods replace the original period of uniform motion. 

Shown in  Fig.~\ref{fig6}  is  the case of movement from Level B1 and Level 4, with pauses and restarts on Levels 1, 2, 3.    It can be seen that when there are  pauses, the period of uniform motion may disappear. In Fig.~\ref{fig6}, the periods with acceleration being $0$ represent pauses. At $0s$,   the elevator starts from Level B1, and accelerates with the acceleration  rapidly increasing. At $1.25s$, the magnitude of the acceleration begins to rapidly decrease, and reaches $0$ at $3.84s$. As the elevator needs to pause on Level 1, there is no period of uniform motion, instead, the acceleration continues to  change, so the sign of the acceleration changes, and the magnitude of the acceleration continues to increase. At $5.12s$, the rate of the change of the acceleration changes, and the acceleration reaches its maximum. After a very short of period of constant acceleration, the acceleration starts to decrease rapidly, and returns to $0$ at about $7.55s$. The antisymmetry between the acceleration and deceleration processes suggests that the velocity returns to 0, when the  elevator pauses on Level 1. During this pause period, there is fluctuation in the acceleration, which is due to the wobbling of the elevator as a result of rapid moving.

Then the elevator restarts on $19.93s$, accelerating upwards, with  its magnitude rapidly increases. The rate of the change of acceleration decreases at 20.55s, and the acceleration reaches its maximum at $21.08s$, followed by a  very short period in which the acceleration remains almost unchanged. At $21.84s$, it starts rapid decrease and returns to $0$ at $22.56s$. Then there is no period of uniform motion, and the acceleration continues to change  in the original rate, and reaches the maximum at $21.72s$; then immediately it starts to decrease, returning to 0 at $27.85s$. The antisymmetry between the accelerating and decelerating periods suggests that the velocity returns to $0$. Now the elevator pauses on Level 2, and there are minor fluctuations in the acceleration during the pausing period.  

Then the elevator restarts on $39.84s$, accelerating upwards, and its magnitude rapidly increases, and  reaches its maximum at $41.28s$, followed by a  very short period in which the acceleration remains almost unchanged. At $41.99s$, it starts rapid decrease and returns to $0$ at $42.93s$. There is no period of uniform motion, and the acceleration continues to change in the original rate, and decreases the magnitude of the rate at $44.34s$. The acceleration  reaches the maximum at $45.16s$; after a very brief period of unchanged acceleration, it starts to decrease rapidly  at $45.69 s$, almost returning to $0$ at $46.1s$. The antisymmetry between the accelerating and decelerating periods suggests that the velocity returns to $0$. Now the elevator pauses on Level 2, and there are minor fluctuations in the acceleration during the pausing period.

Then the elevator restarts on $56.91s$, accelerating upwards, and its magnitude rapidly increases  and  reaches its maximum at $57.59s$, followed by a  very short period in which the acceleration remains almost unchanged. At $59.19s$, it starts rapid decrease and returns to $0$ at $60.03s$. There is no period of uniform motion, and the acceleration continues to change in the original rate, and decreases the magnitude of the rate at $60.82s$,  and reaches the maximum at $61.52s$;  after a very brief period of unchanged acceleration, it starts to decrease rapidly  at $61.99s$,   returning to $0$ at $64.38s$. The antisymmetry between the accelerating and decelerating periods suggests that the velocity returns to $0$. Now the elevator reaches Level  4, the destination.  

It can be seen that for every pause during the movement, an additional process of deceleration-pause-acceleration is inserted. For example,  for the pause on Level  1, the elevator decelerates from $3.84s$ to $7.55s$, pause from $7.55s$ to $19.93s$, and accelerates from $19.93s$ to $22.56s$. 

\section{Conclusion}

The experiment demonstrates that we can conveniently measure the acceleration of the elevator by using the apparent weight inside it. The change of the acceleration of the elevator during its movement is quite stable, with the disturbance and fluctuation minor. 

By comparison, with the initial and final places replaced, the changes of the apparent weight and acceleration  change the signs while the magnitudes remain unchanged. 

From the initial place to the final places, the elevator experiences the accelerating   process (including the increase of the acceleration, the acceleration remains the maximum, and the decrease of the acceleration), the uniform motion, and the decelerating process (  including  the increase of the magnitude of the acceleration, the magnitude of the acceleration remains the maximum, and the magnitude of the acceleration decreases). In the decelerating process, the directions of the acceleration and the velocity are opposite to those in the accelerating process.  When the running distance of the elevator is too short, the period of the uniform motion disappears. 

For every pause during the movement, an additional process of deceleration-pause-acceleration is inserted, replacing the original period of uniform motion. The elevator also decide whether to reduce or cancel the periods of constant acceleration and uniform motions. 

Our work demonstrates a vivid way of illustrating the principles and quantitatively demonstrating the weight loss and overweight, and provides a way of measuring accelerations. For example, a meter of acceleration could be installed in the elevator by using this principle and helps the diagnostic of the fault.  The method can also be used in other cases of rise and decline, e.g. aeroplanes,  rockets, man-made satellites, spaceships, space stations, etc.

\end{document}